\documentclass{JHEP3}
\usepackage{amsmath}
\usepackage{amssymb}
\usepackage{graphicx}
\usepackage{cite}
\usepackage{tabularx}
\usepackage{array}
\usepackage{dcolumn}
\title{Analytic continuation from imaginary to real chemical potential in two-color 
QCD}

\author{Paolo Cea \\
Dipartimento di Fisica, Universit\`a di Bari, and INFN - Sezione di Bari, \\
via Amendola 173, I-70126 Bari, Italy\\
\email{paolo.cea@ba.infn.it}}

\author{Leonardo Cosmai\\
INFN - Sezione di Bari, \\
via Amendola 173, I-70126 Bari, Italy\\
\email{leonardo.cosmai@ba.infn.it}}

\author{Massimo D'Elia\\
Dipartimento di Fisica, Universit\`a di Genova and INFN - Sezione di Genova, \\
via Dodecaneso 33, I-16146 Genova, Italy\\
\email{massimo.delia@ge.infn.it} }

\author{Alessandro Papa\\
Dipartimento di Fisica, Universit\`a della Calabria, and INFN - Gruppo 
collegato di Cosenza, \\
Ponte Bucci, cubo 31C, I-87036 Arcavacata di Rende, Cosenza, Italy\\
\email{papa@cs.infn.it}}

\received{\today}

\abstract{The method of analytic continuation from imaginary to real chemical
potential is one of the most powerful tools to circumvent the sign problem
in lattice QCD. Here we test this method in a theory, two-color QCD, which is
free from the sign problem. We find that the method gives reliable
results, within appropriate ranges of the chemical potential, and that
a considerable improvement can be achieved if suitable functions are
used to interpolate data with imaginary chemical potential.}


\preprint{BARI-TH 2006/557\\GEF-TH-18-06}

\begin{document}

\newcommand{\be}{\begin{equation}}
\newcommand{\ee}{\end{equation}}
\newcommand{\im}{\hat\mu_I}

\section{Introduction}
\label{sec:intro}

The phase diagram of QCD in the temperature -- chemical potential plane is the subject
of many present investigations. Understanding the different phases and the transitions
among them has strong implications in cosmology, in astrophysics and in the phenomenology of
heavy ion collisions. Unfortunately, perturbation theory and approaches based on effective models
can handle a limited number of issues of the QCD phase diagram and, in fact, the lattice formulation
is the only tool for a quantitative approach to the problem based on first principles.
For non-zero chemical potential, however, the QCD fermion determinant becomes complex and the
probability interpretation of the QCD Euclidean action, necessary for the standard Monte Carlo
importance sampling, is lost, this being the well-known ``sign problem''.

Several methods have been invented to circumvent this problem (for a
review, see~\cite{Philipsen:2005mj} and~\cite{Schmidt:2006us}): the
reweighting from the ensemble at
$\mu=0$~\cite{Ferrenberg:1988yz,Ferrenberg:1989ui,Barbour:1997bh,Fodor:2001au},
the Taylor expansion
method~\cite{Allton:2002zi,Gottlieb:1988cq,Choe:2001ar,Choe:2002mt,Choe:2001cq,Gavai:2003mf,Allton:2003vx,
Allton:2005gk,Ejiri:2005uv}, the canonical
approach~\cite{Hasenfratz:1991ax,Alford:1998sd,deForcrand:2006ec,Kratochvila:2005mk},
the density of states
method~\cite{Bhanot:1986ku,Bhanot:1986kv,Karliner:1987cu,Azcoiti:1990ng,Luo:2001id,Gocksch:1988iz,Ambjorn:2002pz}
and the method of analytic continuation from an imaginary chemical
potential~\cite{Lombardo:1999cz,Hart:2000ef,deForcrand:2002ci,deForcrand:2002yi,deForcrand:2003hx,deForcrand:2003ut,D'Elia:2002pj,D'Elia:2002gd,D'Elia:2003uy,Giudice:2004se,Giudice:2004pe,Azcoiti:2004ri,Azcoiti:2004rj,Azcoiti:2004rm,Azcoiti:2005tv,D'Elia:2004at,Chen:2004tb,Kim:2005ck,Lombardo:2005ks,D'Elia:2005qu,Papa:2006jv,Karbstein:2006er}.
Their application has allowed to get relevant information on the
critical line separating the hadronic phase from the quark-gluon
plasma phase in the region $\mu/T\lesssim 1$.

In this paper we focus our attention on the method of analytic
continuation. The idea behind this method is very simple: numerical
simulations are performed at {\it imaginary} chemical potential,
$\mu=i\mu_I$, for which the fermion determinant is real, then Monte
Carlo determinations are interpolated by a suitable function and
finally this function is analytically continued to real values of
$\mu$. This method is rather powerful since the coupling $\beta$ and
the chemical potential $\mu$ can be varied independently and there
is no limitation from increasing lattice size, as happens with other
methods, like those based on reweighting. There is, however, an
important drawback: the periodicity of the QCD partition function
and the presence of non-analyticities arising for imaginary values
of the chemical potential~\cite{Roberge:1986mm} restrict the region
useful for numerical determinations to the strip
$0\leq\mu_I/T<\pi/3$, or even less in presence of ``physical'' phase
transitions. This implies that the accuracy in the interpolation of
the results at imaginary chemical potential has a strong impact on
the extension of the domain of real $\mu$ values reachable after
analytic continuation. In that sense it is very important to answer
the question about which is the optimal way to extract information
from data at imaginary chemical potential, i.e. which is the best
choice for the interpolating function, which only in a some cases
can be guided by physical intuition, leading to some particular
prediction for the behaviour at real $\mu$.
 Moreover one should
always be careful about the actual ranges of applicability of the
method, which can be influenced by the various physical and unphysical
transitions present in the QCD phase diagram, leading to possible non-analyticities.

So far, the method of analytic continuation has been applied in
SU(3) with $n_f=2$~\cite{deForcrand:2002ci,deForcrand:2002yi},
$n_f=3$~\cite{deForcrand:2003hx,deForcrand:2003ut} and
$n_f=4$~\cite{D'Elia:2002pj,D'Elia:2002gd,D'Elia:2003uy,Chen:2004tb}.
Moreover, it has been tested in several theories which do not suffer
from the sign problem, by direct comparison of the analytic
continuation with Monte Carlo results obtained at real
$\mu$~\cite{Hart:2000ef,Giudice:2004se,Giudice:2004pe,Kim:2005ck}.
In most of these applications, a truncated Taylor series (or, more
simply, a polynomial) has been used as interpolating function,
sometimes a Fourier sum for the low temperature region~\cite{D'Elia:2002pj,D'Elia:2002gd}.

The aim of this paper is to study limitations and possible improvements of the method of
analytic continuation, by considering its application to SU(2) or two-color QCD. This theory
is free from the sign problem and Monte Carlo numerical simulations at {\it real} values of
the chemical potential are feasible. This allows to compare the extrapolations
from imaginary to real chemical potential with direct determinations
allowing at the same time both to
discriminate among different Ans\"atze for the interpolating functions
and to directly test the range of reliability of the method itself. The
experience gained in this way can then be hopefully used as a guide
in applications to the real theory.

\FIGURE[ht]{
\includegraphics[width=0.48\textwidth]{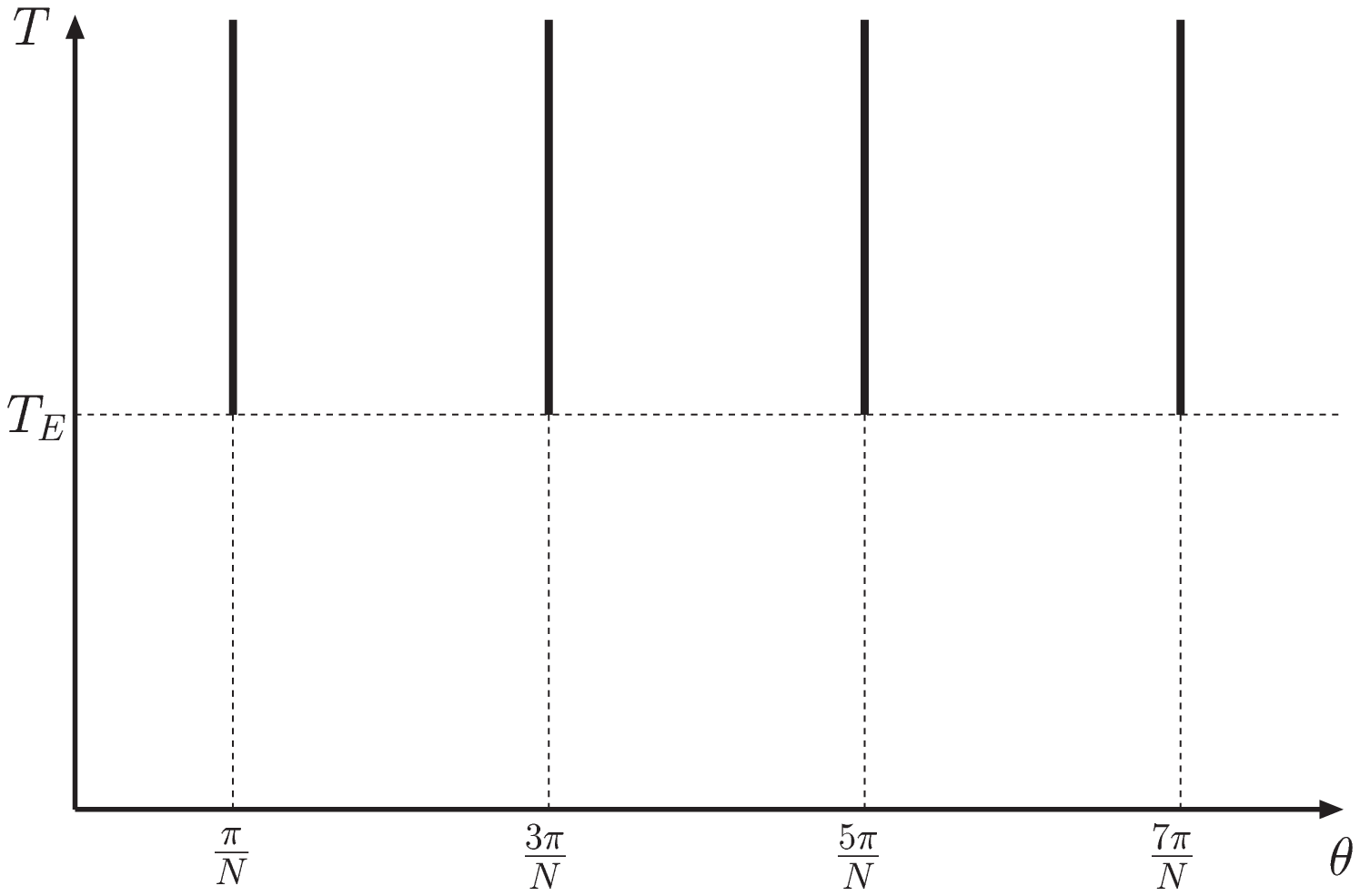}
\includegraphics[width=0.48\textwidth]{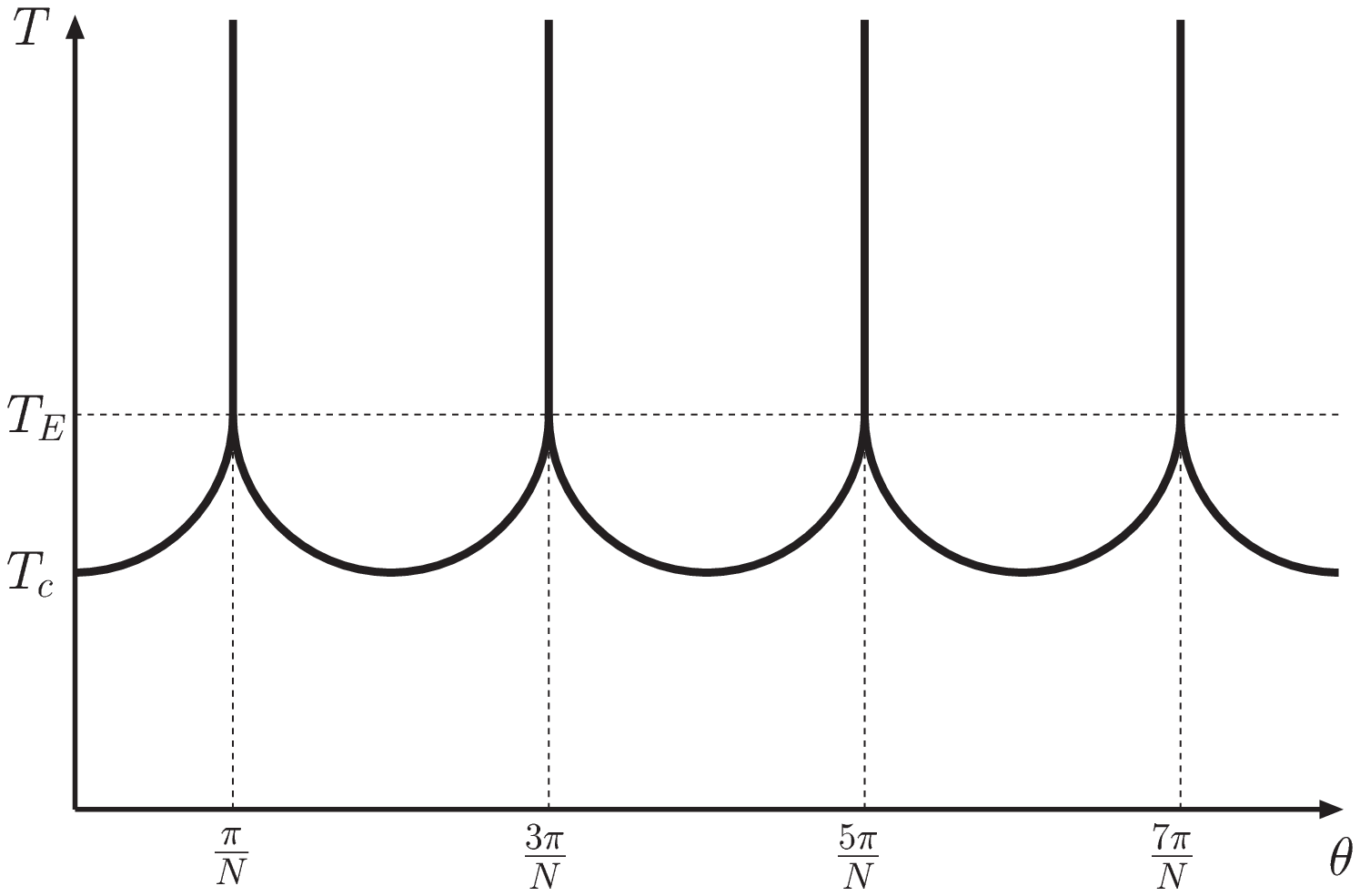}
\caption[]{(Left) Phase diagram in the $(T,\theta)$ plane according
to Ref.~\cite{Roberge:1986mm}. (Right) Tentative phase diagram in
the $(T,\theta)$ plane after the inclusion of the chiral critical
lines.} \label{RW}
}

\FIGURE[ht]{
\includegraphics[width=0.80\textwidth]{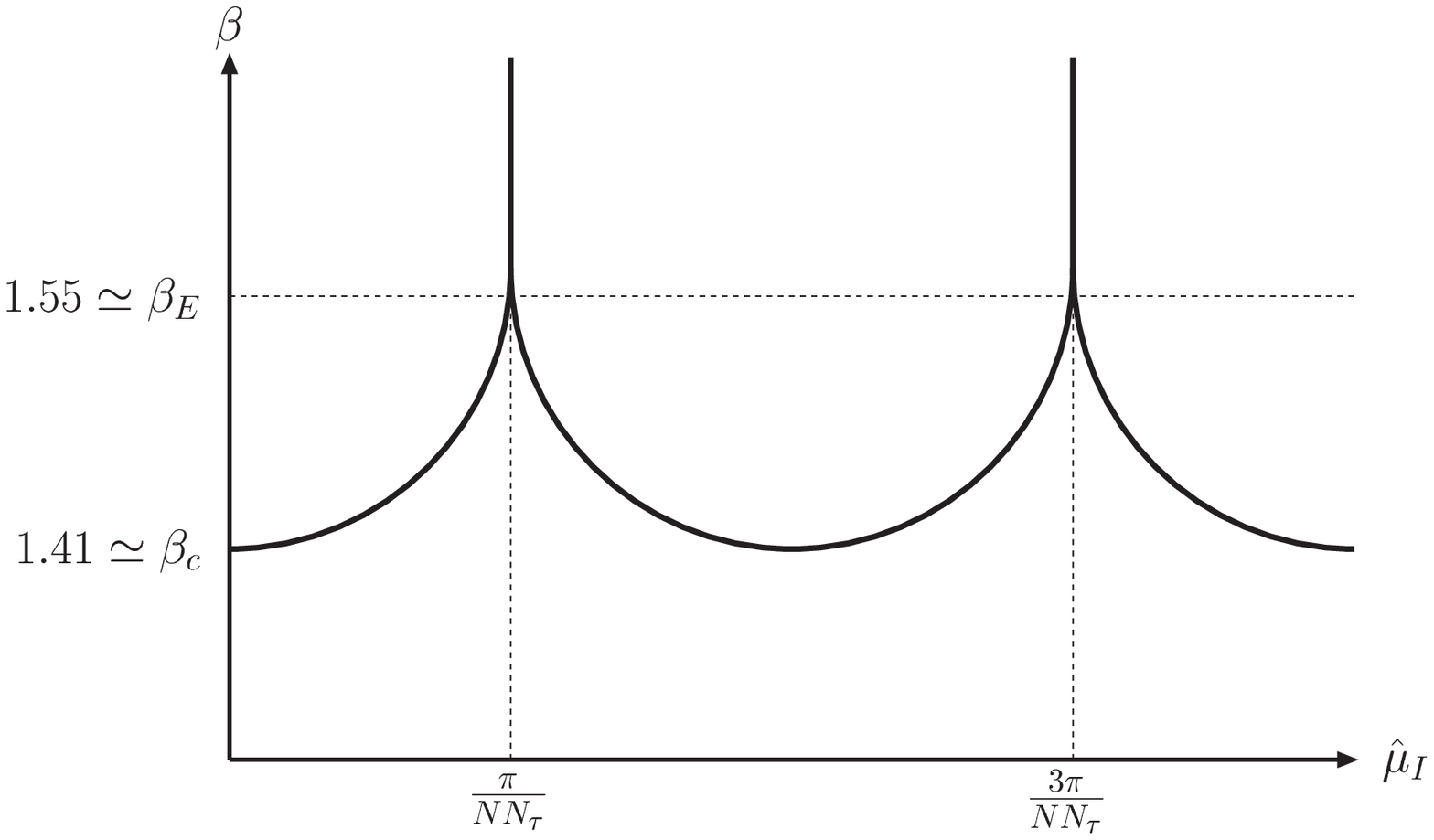}
\caption[]{Phase diagram in the $(\beta,\hat\mu_I)$-plane; $N$ is
the number of colors, $N_\tau$ the extension of the lattice in the
temporal direction. The numerical values for $\beta_E$ and $\beta_c$
are valid for SU(2) in presence of $n_f=8$ degenerate staggered
fermions with mass $am=0.07$.} \label{RWbeta}
}

The paper is organized as follows: in Section~\ref{sec:back}, we briefly recall some general
properties of the phase diagram of SU(N) gauge theories in the temperature - imaginary
chemical potential plane and discuss their implications on the method of analytic
continuation; in Section~\ref{sec:res}, we present our numerical
results and discuss both the choice of the best interpolating
function, showing that functions different from
polynomials can considerably improve the method, and the ranges where
analytic continuation is reliable; finally, in Section~\ref{sec:concl}, we draw our conclusions.

\section{Theoretical background}
\label{sec:back}

Long ago Roberge and Weiss (RW) have shown~\cite{Roberge:1986mm}
that the partition function of any SU(N) gauge theory with non-zero
temperature and imaginary chemical potential, $\mu=i\mu_I$, is
periodic in $\theta\equiv\mu_I/T$ with period $2\pi/N$ and that the
free energy $F$ is a regular function of $\theta$ for $T < T_E$,
while it is discontinuous at $\theta=2\pi(k+1/2)/N$,
$k=0,1,2,\ldots$, for $T > T_E$, where $T_E$ is a characteristic
temperature, depending on the theory. The resulting phase diagram in
the $(T,\theta)$-plane is given in Fig.~\ref{RW} (left), where the
vertical lines represent first order transition lines. This
structure is compatible with the $\mu\to -\mu$ symmetry, related
with CP invariance, and with the Roberge-Weiss periodicity. The
$\mu_I$-dependence of any observable is completely determined if
this observable is known in the strip $0\leq \theta < \pi/N$. It may
be useful to recall the two steps in the proof of periodicity in
SU(N): first, the phase transformation
\begin{equation}
\psi(\vec x,\tau) \longrightarrow \exp(i\tau\mu_I)\psi(\vec x,\tau) \;,
\label{trasnf1}
\end{equation}
then a gauge transformation with periodicity up to an element of the center group Z(N), i.e.
a transformation with gauge group elements $U(\vec x, \tau)$ satisfying the boundary condition
\begin{equation}
U(\vec x,aN_\tau) = \exp(2\pi ik/N) U(\vec x,0) \;, \;\;\;\;\; k \mbox{ integer}\;,
\label{trasnf2}
\end{equation}
where $N_\tau$ is the lattice size in the temporal direction and $a$
is the lattice spacing. The RW periodicity of the partition function
extends to the observables which are left unchanged by these two
transformations. This is certainly the case of the chiral condensate
$\langle \bar \psi \psi \rangle$. The Polyakov loop
$L\equiv\prod_{\tau=1,N_\tau} U(\vec x,a\tau)$, instead, takes the
factor $\exp(2\pi ik/N)$ under the transformation~(\ref{trasnf2}),
which implies that $\langle L \rangle$ moves, continuously or
discontinuously according to the temperature, from one Z(N) sector
to the other when $\mu_I$ passes from one RW sector to the next. As
a consequence, the chiral condensate has the same periodicity in
$\mu_I/T$ of the partition function, $2\pi/N$, while $\langle L
\rangle$ has periodicity in $\mu_I/T$ equal to $2\pi$. These
predictions have been confirmed numerically in several
cases~\cite{deForcrand:2002ci,deForcrand:2002yi,D'Elia:2002pj,D'Elia:2002gd,D'Elia:2003uy,Giudice:2004se,Giudice:2004pe}.

A phase diagram like that in Fig.~\ref{RW} (left) would imply the
absence of any transition along the $T$ axis in the physical regime
of zero chemical potential for any value of $N$, $n_f$ and the quark
masses, which cannot be true. Therefore, it is necessary to admit
that the phase diagram in the $(T,\theta)$-plane is more complicated
than in Fig.~\ref{RW} (left). The simplest possibility is given in
Fig.~\ref{RW} (right), where the added lines generally represent
transitions which can be first order, second order or crossover, and
can be considered as the continuation of the physical critical line
taking place for real chemical potentials. The temperature $T_c$ is
the critical or pseudo-critical one for the transition at zero
chemical potential. The temperature $T_E$ represents the endpoint of
the RW transition lines: the fact that the continuation of the
physical critical line ends right on $T_E$ is not expected a priori,
but is the result of numerical
investigations~\cite{deForcrand:2002ci,deForcrand:2002yi,D'Elia:2002pj,D'Elia:2002gd,D'Elia:2003uy}.

It is convenient to redraw the phase diagram of Fig.~\ref{RW} (right) in the
$(\beta,\hat \mu_I)$-plane (Fig.~\ref{RWbeta}),
where $\beta=2 N/g^2$, $\hat \mu_I\equiv a \mu_I$ is the imaginary chemical potential in lattice units
and it has been used the fact that $T=1/(a N_\tau)$.

Given this phase diagram, it is possible to distinguish three different regimes, corresponding
to different ranges of temperature (i.e. of $\beta$), where analytic
continuation can apply differently.

{\bf Regime a}: $T > T_E$ (or $\beta>\beta_E$).

This regime corresponds to temperatures for which the only expected
non-analyticity at imaginary chemical potential is represented by the RW transition line.
In this case the useful interval in $\hat\mu_I$ for numerical simulations is $[0,\pi/8]$.
On the side of the real chemical potential, no transition line is expected. This situation is,
in some sense, the best possible for the application of the method of analytic continuation.
Simulations at imaginary $\mu$ can be done on a relatively large interval and, if the optimal
interpolating function is found, its continuation should reproduce data for any real value of
$\mu$. The last expectation could actually be wrong if the
critical behaviour induced by the RW line had some influence on the
ranges of analyticity for the partition function also for real values
of the chemical potential: this is an important point that can be
directly checked in two-color QCD.

{\bf Regime b}: $T_c < T < T_E$ (or $\beta_c< \beta< \beta_E$).

This regime corresponds to temperatures for which a non-analyticity is expected at a $\hat \mu_I$ value smaller
than $\pi/8$. On the side of the real chemical potential, no transition line is expected.
This situation is similar to the previous, with the important difference that
the useful interval in $\hat\mu_I$ for numerical simulations is
restricted and the critical behaviour induced by the transition
line may be different, thus making in practice more difficult to find the optimal interpolation.

{\bf Regime c}: $T < T_c$ (or $\beta< \beta_c$).

This regime corresponds to temperatures for which no
non-analyticities should be met in $\hat \mu_I$. This implies that $\hat\mu_I$ can be
varied at will, although no additional information for the observables
of interest here can be gotten by going farther than $\hat\mu_I=\pi/4$, owing to the RW periodicity.
This regime of temperatures is probably the most interesting for physics, since a transition is expected
here for a certain {\it real} value of the chemical potential. This implies that, no matter how good is
the interpolation of data at imaginary chemical potential, its continuation to real $\mu$ should
fail to reproduce data above a certain value.

\FIGURE[ht]{
\includegraphics[width=0.80\textwidth]{final_b190_ndens_new.eps}
\caption[]{Negative side of the horizontal axis: imaginary
part of the fermionic number density {\it vs.} the imaginary
chemical potential at $\beta=1.90$. Positive side of the horizontal
axis: real part of the fermionic number density {\it vs.} the real
chemical potential at $\beta=1.90$.
The green (blue) solid line represents the polynomial (ratio of polynomials)
interpolating function; the dashed lines give the corresponding uncertainty,
coming from the errors in the parameters of the fit.} \label{b190_ndens}
}

\FIGURE[ht]{
\includegraphics[width=0.80\textwidth]{final_b190_psibpsi_new.eps}
\caption[]{Chiral condensate {\it vs.} $\mu^2$ at $\beta=1.90$.
The green (blue) solid line represents the polynomial (ratio of polynomials)
interpolating function; the dashed lines give the corresponding uncertainty,
coming from the errors in the parameters of the fit.} \label{b190_psibpsi}
}

\FIGURE[ht]{
\includegraphics[width=0.80\textwidth]{final_b190_poly_new.eps}
\caption[]{As in Fig.~\ref{b190_psibpsi} for the Polyakov loop.} \label{b190_poly}
}


\FIGURE[ht]{
\includegraphics[width=0.80\textwidth]{chisq_poly190.eps}
\caption[]{$\chi^2$/d.o.f. of the global fit to real and imaginary
chemical potential data for the Polyakov loop at $\beta = 1.90$, for
various fitting functions, as the maximum value $\mu_{\rm max}$ of
real chemical potential taken into account is varied. Lines
connecting data point have been drawn to guide the eye.}
\label{chisq_poly190}
}

\section{Numerical results}
\label{sec:res}

We have performed numerical simulations on a $16^3\times 4$ lattice
of the SU(2) gauge theory with $n_f=8$ degenerate staggered fermions
having mass $am=0.07$. For this theory the tentative phase diagram
looks like in Fig.~\ref{RWbeta}, with $\beta_E\simeq
1.55$~\cite{Giudice:2004se,Giudice:2004pe} and $\beta_c \simeq
1.41$~\cite{Liu:2000in}. The algorithm adopted has been the usual 
exact $\phi$ algorithm described in Ref.~\cite{Gottlieb:1987mq},
properly modified for the inclusion of a finite chemical potential
by multiplying the forward (backward) temporal part of the 
Dirac matrix by $e^{\hat\mu}$ ($e^{-\hat \mu}$), for the case of a real
chemical potential, and by $e^{i \hat\mu_I}$ ($e^{-i \hat\mu_I}$) for the
case of an imaginary chemical potential. In particular that implies,
for real chemical potentials, the impossibility of exploiting the
usual even-odd factorization trick for reducing the number of flavors 
from 8 to 4.
The choice of 8 flavors is therefore linked to the
need of using an exact Hybrid Monte Carlo algorithm: the last is an
unavoidable requirement if we want to make a detailed comparison of
data at imaginary values of $\mu$ with data at real values of $\mu$,
since systematic effects due to an inexact algorithm could be
different for the two cases. The choice of a large volume is instead
essential if we want to make a careful test of the method of
analytic continuation, since possible non-analyticities will show up
only in the thermodynamic limit. The observables we determined are
the Polyakov loop, the chiral condensate and the fermionic number
density $\langle n_q\rangle$.

\FIGURE[ht]{
\includegraphics[width=0.80\textwidth]{final_b145_ndens_new.eps}
\caption[]{As in Fig.~\ref{b190_ndens} for $\beta=1.45$.} \label{b145_ndens}
}

\FIGURE[ht]{
\includegraphics[width=0.80\textwidth]{final_b145_psibpsi_new.eps}
\caption[]{As in Fig.~\ref{b190_psibpsi} for $\beta=1.45$.} \label{b145_psibpsi}
}

\FIGURE[ht]{
\includegraphics[width=0.80\textwidth]{final_b145_poly_new.eps}
\caption[]{As in Fig.~\ref{b190_poly} for $\beta=1.45$.} \label{b145_poly}
}


\FIGURE[ht]{
\includegraphics[width=0.80\textwidth]{chisq_poly145.eps}
\caption[]{As in Fig.~\ref{chisq_poly190} for $\beta=1.45$.}
\label{chisq_poly145}
}

We have considered three $\beta$ values, $\beta$=1.90, 1.45 and
1.30, corresponding to the three different regimes exposed in
Section~\ref{sec:back}, and for each we have taken measurements for
several values of the chemical potential, both imaginary and real.
The summary of numerical simulations is given in
Tables~\ref{summ_murea_b190}, \ref{summ_muimm_b190},
\ref{summ_murea_b145}, \ref{summ_muimm_b145}, \ref{summ_murea_b130},
\ref{summ_muimm_b130}. Simulations have been performed on the APE100
and APEmille crates in Bari and on the recently installed computer
facilities at the INFN apeNEXT Computing Center in Rome. Statistics
have been chosen so as to have statistical errors well below 1\% in
most cases: indeed our ability to discern the best among a set of
possible interpolating functions as well as to detect the exact
ranges beyond which analytic continuation fails, is strictly related
to the statistical precision of our data.

We have chosen two different strategies for our analysis.
We have used the data at imaginary chemical potential $\mu_I$ to determine the parameters of the
interpolating function, then we have analytically continued this function to real values of the
chemical potential and compared there with direct Monte Carlo
determinations. In this way we are able to test how the method of analytic
continuation is able to reproduce the correct physical results for
real values of $\mu$, and to understand which is the best suited
function to do so.

As an alternative way to analyze our results,
we have tried to fit both sets of data together, at imaginary
and real chemical potential, with several analytic functions and
using variable ranges for both $\hat\mu$ and $\im$.
In this way, using the $\chi^2$ test as a statistical tool,
we are able to understand in which ranges, if any, the method of analytic continuation
makes any sense at all, at least within the
set of analytic functions taken into considerations.

In order to fulfill
CP invariance, the interpolating function must be a even function of $\mu$ for observables, such as the
Polyakov loop and the chiral condensate, which do not depend explicitly on $\mu$. The fermionic
number density, being the logarithmic derivative of the partition function with respect to the
chemical potential, is instead an odd function of $\mu$.

\FIGURE[ht]{
\includegraphics[width=0.80\textwidth]{final_b130_ndens.eps}
\caption[]{Negative side of the horizontal axis: imaginary
part of the fermionic number density {\it vs.} the imaginary
chemical potential at $\beta=1.30$. Positive side of the horizontal
axis: real part of the fermionic number density {\it vs.} the real
chemical potential at $\beta=1.30$. The green (blue) solid line represents
the Fourier (polynomial) interpolating function and
its continuation; the dashed lines give the corresponding uncertainty,
coming from the errors in the parameters of the fit.} \label{b130_ndens}
}

\FIGURE[ht]{
\includegraphics[width=0.80\textwidth]{final_b130_psibpsi.eps}
\caption[]{Chiral condensate {\it vs.} $\mu^2$ at $\beta=1.30$.
The green (blue) solid line represents the Fourier (polynomial) 
interpolating function and its continuation; the dashed lines give 
the corresponding uncertainty, coming from the errors in the 
parameters of the fit.} \label{b130_psibpsi}
}

\FIGURE[ht]{
\includegraphics[width=0.80\textwidth]{final_b130_poly.eps}
\caption[]{As in Fig.~\ref{b130_psibpsi} for the Polyakov
loop.} \label{b130_poly}
}


\FIGURE[ht]{
\includegraphics[width=0.80\textwidth]{chisq_130.eps}
\caption[]{$\chi^2$/d.o.f. of the global fit to real and imaginary
chemical potential data for the various observable at $\beta =
1.30$, as a function of the maximum value $\mu_{\rm max}$ of real
chemical potential taken into account. Fitting functions are the
same used for Figs.~\ref{b130_ndens}, \ref{b130_psibpsi}, \ref{b130_poly}
and reported in Eq.~(\ref{fourier_cos8}), (\ref{fourier_cos4}) and
(\ref{fourier_sin}). The values of $\chi^2$/d.o.f. are quite stable
around one for all observables, till $\hat\mu_{\rm max}$ crosses the
pseudo-critical point.} \label{chisq_130}
}

\FIGURE[ht]{
\includegraphics[width=0.80\textwidth]{chiral_susc_b130.eps}
\caption[]{Susceptibility of the chiral condensate at $\beta=1.30$
for real $\hat\mu$.} \label{chiral_susc_b130}
}

We separate the discussion of our results for the three different
regimes, reflecting the different strategies
followed in searching for the optimal interpolation and the different
behaviors observed for the physical observables.

\subsection{The high temperature region $\beta > \beta_E$}

For this region we have used two kinds of interpolating functions
for the data at imaginary $\mu$: polynomials and ratio of polynomials,
the last choice being related to the use of  Pad\'e
approximants suggested in Ref.~\cite{Lombardo:2005ks}.
For the Polyakov loop and the chiral condensate we have considered a second order polynomial in $\mu^2$,
\begin{equation}
A+B \hat\mu_I^2+C \hat\mu_I^4\;,
\label{poly}
\end{equation}
according to the standard approach, and the ratio of two first order polynomials in $\mu^2$,
\begin{equation}
\frac{A+B \hat\mu_I^2}{1+C \hat\mu_I^2}\;,
\label{ratio}
\end{equation}
according to our new proposal.
Similarly, for the fermionic number density we have used a polynomial of the form
\begin{equation}
A \hat\mu_I+B \hat\mu_I^3+C \hat\mu_I^5\;,
\label{poly_ndens}
\end{equation}
and the ratio
\begin{equation}
\frac{A\hat\mu_I+B \hat\mu_I^3}{1+C \hat\mu_I^2}\;.
\label{ratio_ndens}
\end{equation}

Our findings at $\beta=1.90$ are summarized in Figs.~\ref{b190_ndens}, \ref{b190_psibpsi},
\ref{b190_poly} and Table~\ref{param_b190}.
In Fig.~\ref{b190_ndens} we put on the same plot the imaginary part of the fermionic number density
as a function of $\mu_I$ and the real part of the fermionic number density as a function of the
real $\mu$. The two data sets match smoothly at $\mu=0$, which is a necessary condition for the
applicability of the method of analytic continuation. The fermionic density approaches two for
large values of the real chemical potential. This saturation effect is
a lattice artifact, which is due to the
fact that no more than two fermions per site can be accommodated on
the lattice (``Pauli blocking''),
and manifests itself for values of the chemical potential which are
close to the ultraviolet cutoff, i.e. for $\hat\mu$ of order 1
(see Refs.\cite{Alles:2006ea,Hands:2006ve} for a recent discussion of this phenomenon).
The solid lines represent the two kinds of interpolating functions, whose parameters are determined
by a fit on the data at imaginary chemical potential.
Here both interpolations, polynomial and rational function, nicely reproduce the data
at real $\mu$ over a large interval. Deviations start at values of $\mu$ for which
the saturation effects are certainly important.

In Fig.~\ref{b190_psibpsi} we show the chiral condensate as a function of $\mu^2$.
Again data at imaginary
$\mu$, i.e. $\mu^2<0$, and data at real $\mu$, i.e. $\mu^2>0$, nicely match at $\mu=0$. This time
the different behavior of the two kinds of interpolation clearly emerges. The ratio of first order
polynomials in $\mu^2$ reproduces the data at real $\mu$ on a much larger interval than the second
order polynomial in $\mu^2$. Deviations arise for values of real $\mu$ for which saturation effects
are probably already important.
The same conclusions can be drawn from Fig.~\ref{b190_poly} which shows data and interpolations
for the Polyakov loop.

The above conclusions do not change if larger order terms are included in the polynomial
interpolation~(\ref{poly}). In fact, larger order polynomials fail to reproduce the data at
real $\mu$ even earlier in $\mu$ than second order polynomials. This is due to the fact that the
higher order terms of the polynomial are the less accurately determined in the fit
to data at imaginary $\mu$.  On the other side, if in the ratio of polynomials
the order of the polynomials at the numerator and/or at the denominator is
increased, no improvement is observed.


It is interesting to compare the parameters in the expansion~(\ref{poly_ndens}) with those predicted by the
perturbation theory in $\mu/T$ of the fermionic density. In the infinite temperature limit, and taking
into account the rotation to real chemical potential in Eq.~(\ref{poly_ndens}), it should be $A=-n_f/N_\tau^2=-1/2$
and $B=n_f/\pi^2=8/\pi^2=0.810569...$~\cite{Allton:2003vx}. These values are in rough agreement with our
findings at the largest available $\beta$ (see Table~\ref{param_b190}).

We now expose the results of our combined fits using both sets of data
at imaginary and real values of $\mu$.
The range of values used in the fit is limited on the imaginary
chemical potential side by the presence of the RW transition,
so that we included all data with $\im < \pi/8$. For real chemical
potentials, we have considered the possible presence of
non-analyticities and have repeated our fits for
different values of the maximum real chemical potential, $\hat\mu_{\rm
  max}$.

We report only results obtained for the Polyakov loop: those obtained for the
other two observables look very similar. We have tried fits with
several analytic functions, only a few of them being exemplified in
Fig.~\ref{chisq_poly190},
where we report the value of $\chi^2$/d.o.f. as a function of
$\hat\mu_{\rm max}$.
The outcome of our analysis, as evident from Fig.~\ref{chisq_poly190},
 can be summarized as follows:
acceptable values of $\chi^2$/d.o.f.
are obtained once sufficiently higher order polynomials or ratio of polynomials
are taken into account, but only if $\hat\mu_{\rm max}$
is less than about 0.5.
Instead the value of $\chi^2$/d.o.f. gets sensibly different from one
for larger values of $\hat\mu_{\rm max} \to 1$, regardless of the interpolating function.

We interpret this result as a proof that, within statistical errors, data
at real and imaginary chemical potential can indeed be described by
one only analytic function, even if in a limited range.
Hints of possible non-analyticities appear for real chemical potentials $\hat \mu > 0.5$.
We believe that the most plausible explanation of them is the onset of
saturation effects.

A comment is in order about the use of higher order polynomials.
Fig.~\ref{chisq_poly190} could give the impression that increasing the order
of the polynomial sensibly improves the method of analytic
continuation, since a reasonable  $\chi^2$/d.o.f. is obtained
for a wider range. This could seem to contradict our previous
statements about the choice of the optimal function for extrapolating
data from imaginary values of $\mu$, in fact it is not so.
Indeed, one should consider that when trying to extrapolate
information to $\mu^2 >0$ having at disposal only
information from negative values of $\mu^2$, the use of polynomials in
$\mu^2$ can result in instabilities in the determination of the
coefficients, since a polynomial with positive coefficients for
$\mu^2 < 0$ is continued to a polynomial with alternating
coefficients for $\mu^2 > 0$ and vice versa. These instabilities
clearly disappear if data on both sides are available, but of course this
situation cannot be reproduced for real QCD.

\subsection{The intermediate region $\beta_c < \beta < \beta_E$}

Our findings at $\beta=1.45$ are summarized in Figs.~\ref{b145_ndens}, \ref{b145_psibpsi},
\ref{b145_poly} and Table~\ref{param_b145}.
Also in this case we have used  polynomials and ratio of polynomials
as interpolating functions for the data at imaginary $\mu$.
Here the discussion goes along the same lines as for $\beta=1.90$ with one important
difference: both in the case of polynomials or ratios of polynomials as interpolating functions,
and for any of the three observables considered here, a fit at imaginary chemical potential
with $\chi^2$/d.o.f of the order of one is possible only in the interval $[0,\bar\mu_I]$, with
$\bar\mu_I = 0.22 \div 0.24$. This $\hat\mu_I$ represents the onset of a transition,
which shows up also as a peak in the chiral condensate susceptibility, centered around that value
of $\hat\mu_I$. The fact that the interval in $\hat\mu_I$ available for numerical simulations
is shorter makes the interpolation, and consequently its continuation, less accurate. Nevertheless,
also at this $\beta$, the use of ratio of polynomial performs much better than simple polynomials.

Also in this regime we have performed combined fits using both sets of data
at imaginary and real values of $\mu$.
In this case  the range of values used in the fit is limited, on the imaginary
chemical potential side, by the presence of the continuation of the
physical critical line, so that we have included only values  $\im < 0.20$.

Also in this case we report only results obtained for the Polyakov
loop with several fitting functions, as  exemplified in
Fig.~\ref{chisq_poly145}, as a function of $\hat\mu_{\rm max}$.
Results look very similar to those obtained for $\beta = 1.90$,
with a few differences: while the quality of the fits obtained
with the ratio of polynomials does not change with respect to $\beta = 1.90$,
higher order polynomials are necessary to obtain reasonable
$\chi^2$/d.o.f., and hints of non-analyticities show up generally
earlier, as a function of $\hat\mu_{\rm max}$, than for $\beta =
1.90$. A possible explanation for the different behaviour could reside
in the presence of the physical pseudo-critical line for imaginary values
of $\mu$.

\subsection{The low temperature region $\beta < \beta_c$}

Below $\beta_c$ our observables are smooth functions of $\hat\mu_I$, with periodicity in $\hat\mu_I$
equal to $\pi/4$ in the case of the fermionic number density and of the chiral condensate and equal
to $\pi/2$ in the case of the Polyakov loop. This leads naturally to the use of Fourier sums
as interpolating functions and, in particular,
\begin{equation}
A+B\cos(8\hat\mu_I)+C\cos(16\hat\mu_I)
\label{fourier_cos8}
\end{equation}
for the chiral condensate,
\begin{equation}
A\cos(4\hat\mu_I)+B\cos(12\hat\mu_I)
\label{fourier_cos4}
\end{equation}
for the Polyakov loop and
\begin{equation}
A\sin(8\hat\mu_I)+B\sin(16\hat\mu_I)
\label{fourier_sin}
\end{equation}
for the fermionic number density, which is odd in $\mu_I$.

We summarize our results at $\beta=1.30$ in Figs.~\ref{b130_ndens}, \ref{b130_psibpsi},
\ref{b130_poly} and Table~\ref{param_b130}. The functions chosen for our fits, and reported
in Eq.~(\ref{fourier_cos8}), (\ref{fourier_cos4}) and
(\ref{fourier_sin}), are those containing the minimum number of terms
necessary to obtain a $\chi^2$/d.o.f. close to one (the use of less
terms leading to a sensible increase of $\chi^2$/d.o.f.).
From Table~\ref{param_b130} it is possible to see that the coefficients of the secondary
harmonic terms in the Fourier sums~(\ref{fourier_cos8}), (\ref{fourier_cos4}) and
(\ref{fourier_sin}) are suppressed by a factor of a few tens with respect to the coefficients
of the dominant harmonic, thus signaling a quite fast convergence of the Fourier sums.
It is interesting to notice that the term proportional  to
$\cos(8\hat\mu_I)$ does not appear in Eq.~(\ref{fourier_cos4}),
i.e. one term in the harmonic series for the Polyakov loop
seems to be missing. The reason is that in the low temperature region
center symmetry constrains the Polyakov loop to be zero at $\im = \pi/8$
(corresponding to the border between the two center sectors),
so that all frequencies which are even multiples of $4\hat\mu_I$
must be excluded. We have verified that if these frequencies are included in
the interpolating function, the corresponding coefficients are put to zero by the fit.
The Fourier sums become sums of hyperbolic sine and cosine functions after continuation
to real $\mu$, which diverge very rapidly and reproduce only partially the data at real $\mu$.
The deviation between the extrapolation and the data can be taken as an estimate of the
pseudo-critical value of $\hat\mu_R$. This is confirmed by the study of the chiral condensate
susceptibility, which exhibits a peak centered around that value.

Alternative attempts with longer Fourier sums or with ratios of Fourier sums did not change
this scenario. In the case of polynomials as interpolating functions, the 
behavior is similar to Fourier sums if, however, large order polynomials are used (see blue lines in 
Figs.~\ref{b130_ndens}, \ref{b130_psibpsi}, \ref{b130_poly}). In this case, the interpolation 
of data at imaginary $\mu$ works in an interval shorter than that for Fourier sums. 
These observations confirm that Fourier sums are indeed the natural functions to be used for 
the analytic continuation in the low temperature region.

Regarding the global  combined fits using both sets of data
at imaginary and real values of $\mu$, the results of our analysis
are reported in Fig.~\ref{chisq_130}. In this case we report results
for all observables, with the same
fitting functions, Eq.~(\ref{fourier_cos8}), (\ref{fourier_cos4}) and
(\ref{fourier_sin}), used previously. All data at imaginary chemical
potential are taken into account, since no phase transition at all
is expected on that side, while data at real $\mu$ are limited to a
maximum value $\hat\mu_{\rm max}$.
It clearly emerges that, for all observables, both sets of data can be
nicely fitted by a common analytic function, till $\hat\mu_{\rm max}$
reaches the region where the physical pseudo-critical point is
located;
at that point the method of analytic continuation clearly loses any
sense. However, it is quite interesting to notice that the analytic
properties of the partition function for imaginary chemical potentials
are not influenced at all by the presence of the pseudo-critical point
at real $\mu$.

An independent determination of the pseudo-critical chemical potential
can be obtained
by the study of the susceptibility of the chiral condensate for real
$\hat \mu$, shown in Fig.~\ref{chiral_susc_b130}. There is an
evident peak at $\hat\mu\simeq 0.28$, in good agreement with the
determinations from the $\chi^2$ test method.

\section{Conclusions and outlook}
\label{sec:concl}

We have studied the method of analytic continuation in a theory which does not suffer
from the sign problem and have looked for better interpolating functions at imaginary
$\mu$, to be used instead of the polynomial, as has been done in most cases so far in the literature.

We have verified that data at real and imaginary chemical potential
can indeed be well described by common suitable analytic  functions,
in appropriate ranges, and we have found that a considerable
improvement can be achieved, when extrapolating data from imaginary
to real chemical potentials, if ratios of polynomials (or equivalently Pad\'e
approximants~\cite{Lombardo:2005ks}), are used
as interpolating functions, if the temperature is larger than the pseudo-critical one
at zero chemical potential. Below that value, instead, Fourier sums
seem to be the best Ansatz, as expected and tested also in other contexts~\cite{D'Elia:2002pj,D'Elia:2002gd}.

The deviations from analyticity and between the extrapolated functions and the data at real chemical potential
have different explanations, according to the temperature regime. Above the temperature of
the RW endpoint they arise most likely from unphysical saturation
effects, due to the lattice discretization (``Pauli blocking'').
In the intermediate regime, deviations stem also from the limited range of the interval
in the imaginary chemical potential for the numerical simulations, which makes the interpolation
less easy: this is caused by the presence of a pseudo-critical point
for imaginary values of the chemical potential, which could also
contribute to restrict the range where analytic continuation can be applied.
Finally, in the low temperature regime, deviations necessarily appear in correspondence of the
transition at real chemical potential.

The lessons we have learned from this study and which could be applied to the physically
interesting case of SU(3) can be summarized as follows:
\begin{itemize}

\item above the pseudo-critical temperature, ratio of polynomials should be used as interpolating functions
instead of polynomials; their continuation to real chemical potentials is the more reliable the larger
is the interval of imaginary chemical potential where they succeed in interpolating data;

\item below the pseudo-critical temperature, one should surely use
Fourier sums: they nicely reproduce
data at imaginary chemical potentials, but are extrapolated to hyperbolic functions which rapidly diverge
at real chemical potentials; nevertheless, analytic continuation works
fairly well till the pseudo-critical value of the real chemical potential is reached.

\end{itemize}

\section*{Acknowledgments}

We would like to thank Ph.~de Forcrand, M.P.~Lombardo and O.~Philipsen
for very useful comments and discussions.


\providecommand{\href}[2]{#2}\begingroup\raggedright\endgroup

\newpage

\TABLE[ht]{
\setlength{\tabcolsep}{0.9pc}
\centering
\caption[]{Summary of the simulations at $\beta=1.90$ and real chemical potential,
$\hat\mu=\hat\mu_R$.}
\begin{tabular}{lcclll}
\hline
\hline
  \multicolumn{1}{c}{$\hat\mu_R$} & machine & stat. & \multicolumn{1}{c}{$\langle L \rangle$}
& \multicolumn{1}{c}{$\langle\overline\psi\psi\rangle$}
& \multicolumn{1}{c}{$\Re \langle n_q\rangle$} \\
\hline
0.   & APE100   & 1k & 0.39712(28) & 0.085768(65)  &  0.00048(21)  \\
0.05 & APEmille & 1k & 0.39867(26) & 0.085421(55)  &  0.01907(23)  \\
0.10 & APEmille & 1k & 0.40123(25) & 0.084650(66)  &  0.03878(23)  \\
0.15 & APEmille & 5k & 0.40559(13) & 0.083352(25)  &  0.06030(14)  \\
0.20 & APEmille & 5k & 0.41159(16) & 0.081532(37)  &  0.08413(13)  \\
0.25 & APEmille & 5k & 0.41876(14) & 0.079271(28)  &  0.11133(18)  \\
0.30 & APEmille & 5k & 0.42731(15) & 0.076431(45)  &  0.14280(11)  \\
0.35 & APEmille & 5k & 0.43669(13) & 0.073201(33)  &  0.17937(12)  \\
0.40 & APEmille & 5k & 0.44649(12) & 0.069646(34)  &  0.22207(11)  \\
0.45 & APEmille & 5k & 0.45687(17) & 0.065784(43)  &  0.27166(14)  \\
0.50 & APEmille & 5k & 0.46675(13) & 0.061474(42)  &  0.32895(21)  \\
0.55 & APEmille & 5k & 0.47620(11) & 0.057068(67)  &  0.39490(20)  \\
0.60 & APEmille & 5k & 0.48516(10) & 0.052518(59)  &  0.46992(16)  \\
0.65 & APEmille & 5k & 0.49244(11) & 0.047807(53)  &  0.55385(14)  \\
0.70 & APEmille & 5k & 0.49809(14) & 0.043017(47)  &  0.64675(16)  \\
0.75 & APEmille & 5k & 0.50144(12) & 0.038366(53)  &  0.74778(17)  \\
0.80 & APEmille & 5k & 0.50207(11) & 0.033650(54)  &  0.85613(18)  \\
0.90 & apeNEXT  & 7k & 0.49331(10) & 0.025022(29)  &  1.087610(89) \\
1.00 & apeNEXT  & 7k & 0.46667(11) & 0.017377(20)  &  1.32517(11)  \\
1.10 & apeNEXT  & 7k & 0.41818(17) & 0.011103(28)  &  1.54686(20)  \\
1.20 & apeNEXT  & 7k & 0.34646(23) & 0.006400(17)  &  1.72969(18)  \\
1.50 & apeNEXT  & 7k & 0.13191(24) & 0.000750(12)  &  1.966760(93) \\
1.80 & apeNEXT  & 7k & 0.04066(33) & 0.0000712(93) &  1.996934(99) \\
2.10 & apeNEXT  & 7k & 0.01254(44) & 0.0000033(79) &  1.99972(10)  \\
\hline
\hline
\end{tabular}
\label{summ_murea_b190}
}

\TABLE[ht]{
\setlength{\tabcolsep}{0.9pc}
\centering \caption[]{Summary of the simulations at $\beta=1.90$ and
imaginary chemical potential, $\hat\mu=i\hat\mu_I$.}
\begin{tabular}{lccrlll}
\hline
\hline
\multicolumn{1}{c}{$\hat\mu_I$} & machine & stat. & \multicolumn{1}{c}{$\langle L \rangle$}
& \multicolumn{1}{c}{$\langle\overline\psi\psi\rangle$}
& \multicolumn{1}{c}{$\Im \langle n_q\rangle$} \\
\hline
0.    & APE100  & 1k &    0.39712(28) & 0.085768(65) &    0.00048(21)  \\
0.05  & APE100  & 1k &    0.39636(27) & 0.086028(39) & \hspace{-0.30cm}$-$0.01896(27)  \\
0.075 & apeNEXT & 5k &    0.39525(17) & 0.086394(21) & \hspace{-0.30cm}$-$0.027905(55) \\
0.10  & APE100  & 1k &    0.39323(31) & 0.086856(43) & \hspace{-0.30cm}$-$0.03638(21)  \\
0.125 & apeNEXT & 5k &    0.39133(24) & 0.087473(34) & \hspace{-0.30cm}$-$0.045120(54) \\
0.15  & APE100  & 1k &    0.38855(26) & 0.088213(42) & \hspace{-0.30cm}$-$0.05262(25)  \\
0.175 & apeNEXT & 5k &    0.38518(18) & 0.089108(21) & \hspace{-0.30cm}$-$0.060482(45) \\
0.20  & APE100  & 1k &    0.38159(28) & 0.090086(51) & \hspace{-0.30cm}$-$0.06709(25)  \\
0.225 & apeNEXT & 5k &    0.37717(16) & 0.091346(27) & \hspace{-0.30cm}$-$0.073076(60) \\
0.25  & APE100  & 1k &    0.37185(28) & 0.092688(62) & \hspace{-0.30cm}$-$0.07808(22)  \\
0.275 & apeNEXT & 5k &    0.36596(23) & 0.094159(40) & \hspace{-0.30cm}$-$0.082205(66) \\
0.30  & APE100  & 5k &    0.35986(45) & 0.095916(28) & \hspace{-0.30cm}$-$0.08535(12)  \\
0.325 & apeNEXT & 5k &    0.35893(23) & 0.097733(27) & \hspace{-0.30cm}$-$0.087182(62) \\
0.35  & APE100  & 1k &    0.34173(44) & 0.099686(59) & \hspace{-0.30cm}$-$0.08787(23)  \\
0.375 & apeNEXT & 5k &    0.32987(70) & 0.101984(48) & \hspace{-0.30cm}$-$0.08732(16)  \\
0.40  & APE100  & 1k & $-$0.32419(49) & 0.103085(88) &    0.08697(26)  \\
\hline
\hline
\end{tabular}
\label{summ_muimm_b190}
}

\TABLE[ht]{
\setlength{\tabcolsep}{0.9pc}
\centering \caption[]{Summary of the simulations at $\beta=1.45$ and
real chemical potential, $\hat\mu=\hat\mu_R$.}
\begin{tabular}{lcclll}
\hline
\hline
\multicolumn{1}{c}{$\hat\mu_R$} & machine & stat. & \multicolumn{1}{c}{$\langle L \rangle$}
& \multicolumn{1}{c}{$\langle\overline\psi\psi\rangle$}
& \multicolumn{1}{c}{$\Re \langle n_q\rangle$} \\
\hline
0.   & apeNEXT  & 5k & 0.25078(62) & 0.20791(43)  & \hspace{-0.30cm}$-$0.000013(73) \\
0.   & APEmille & 5k & 0.25058(40) & 0.20786(34)  &  0.00012(22) \\
0.05 & APEmille & 5k & 0.25389(30) & 0.20473(33)  &  0.01889(16) \\
0.10 & APEmille & 5k & 0.26164(27) & 0.19773(32)  &  0.03894(17) \\
0.15 & APEmille & 5k & 0.27373(46) & 0.18706(44)  &  0.06358(16) \\
0.20 & APEmille & 5k & 0.28786(29) & 0.17435(38)  &  0.09185(18) \\
0.25 & APEmille & 5k & 0.30307(27) & 0.16074(20)  &  0.12650(18) \\
0.30 & APEmille & 5k & 0.31919(17) & 0.14730(13)  &  0.16839(26) \\
0.35 & APEmille & 5k & 0.33446(18) & 0.13388(11)  &  0.21787(18) \\
0.40 & APEmille & 5k & 0.34964(16) & 0.120640(87) &  0.27646(19) \\
0.45 & APEmille & 5k & 0.36357(17) & 0.10762(11)  &  0.34454(20) \\
0.50 & APEmille & 5k & 0.37576(15) & 0.09535(10)  &  0.42341(25) \\
0.55 & APEmille & 5k & 0.38624(11) & 0.083452(87) &  0.51294(23) \\
0.60 & APEmille & 5k & 0.39362(14) & 0.07217(12)  &  0.61217(29) \\
0.65 & APEmille & 5k & 0.39781(14) & 0.06176(11)  &  0.72067(24) \\
0.70 & apeNEXT  & 5k & 0.39851(14) & 0.052110(38) &  0.83731(15) \\
0.75 & apeNEXT  & 5k & 0.39481(16) & 0.043325(38) &  0.96042(14) \\
0.80 & apeNEXT  & 5k & 0.38622(12) & 0.035453(54) &  1.08808(17) \\
0.90 & apeNEXT  & 5k & 0.35369(17) & 0.022487(44) &  1.34492(17) \\
1.00 & apeNEXT  & 5k & 0.30183(12) & 0.013127(33) &  1.57817(15) \\
1.10 & apeNEXT  & 5k & 0.23811(15) & 0.006995(26) &  1.75856(14) \\
1.20 & apeNEXT  & 5k & 0.17580(20) & 0.003483(22) &  1.87501(10) \\
1.50 & apeNEXT  & 5k & 0.05762(15) & 0.000345(19) &  1.98708(12) \\
1.80 & apeNEXT  & 5k & 0.01749(15) & 0.000046(10) &  1.998941(87)\\
2.10 & apeNEXT  & 5k & 0.00535(14) & 0.0000101(81)&  1.999779(80)\\
\hline
\hline
\end{tabular}
\label{summ_murea_b145}
}

\TABLE[ht]{
\setlength{\tabcolsep}{0.9pc}
\centering \caption[]{Summary of the simulations at $\beta=1.45$ and
imaginary chemical potential, $\hat\mu=i\hat\mu_I$.}
\begin{tabular}{lcclll}
\hline
\hline
\multicolumn{1}{c}{$\hat\mu_I$} & machine & stat. & \multicolumn{1}{c}{$\langle L \rangle$}
& \multicolumn{1}{c}{$\langle\overline\psi\psi\rangle$}
& \multicolumn{1}{c}{$\Im \langle n_q \rangle$} \\
\hline
0.   & apeNEXT  & 5k & 0.25078(62) & 0.20791(43) & \hspace{-0.30cm}$-$0.000013(73) \\
0.   & APEmille & 5k & 0.25058(40) & 0.20786(34) &    0.00012(22)  \\
0.02 & apeNEXT  & 5k & 0.25026(35) & 0.20811(37) & \hspace{-0.30cm}$-$0.007369(67) \\
0.04 & apeNEXT  & 5k & 0.24890(39) & 0.20925(54) & \hspace{-0.30cm}$-$0.01450(11)  \\
0.05 & APE100   & 1k & 0.24811(59) & 0.21009(54) & \hspace{-0.30cm}$-$0.01738(42)  \\
0.06 & apeNEXT  & 5k & 0.24647(36) & 0.21156(41) & \hspace{-0.30cm}$-$0.021307(83) \\
0.08 & apeNEXT  & 5k & 0.24200(43) & 0.21543(36) & \hspace{-0.30cm}$-$0.027878(84) \\
0.10 & apeNEXT  & 5k & 0.23764(60) & 0.21958(60) & \hspace{-0.30cm}$-$0.033936(80) \\
0.10 & APE100   & 1k & 0.23772(50) & 0.21962(74) & \hspace{-0.30cm}$-$0.03346(38)  \\
0.12 & apeNEXT  & 5k & 0.23050(47) & 0.22575(60) & \hspace{-0.30cm}$-$0.039025(92) \\
0.14 & apeNEXT  & 5k & 0.22363(49) & 0.23198(46) & \hspace{-0.30cm}$-$0.04344(14)  \\
0.15 & APE100   & 1k & 0.21874(68) & 0.23556(65) & \hspace{-0.30cm}$-$0.04552(44)  \\
0.16 & apeNEXT  & 5k & 0.21316(63) & 0.24050(65) & \hspace{-0.30cm}$-$0.04637(19)  \\
0.18 & apeNEXT  & 5k & 0.19988(62) & 0.25118(64) & \hspace{-0.30cm}$-$0.04788(15)  \\
0.20 & apeNEXT  & 5k & 0.18395(93) & 0.26395(79) & \hspace{-0.30cm}$-$0.04787(23)  \\
0.20 & APE100   & 5k & 0.18626(61) & 0.26207(43) & \hspace{-0.30cm}$-$0.04860(27)  \\
0.22 & apeNEXT  & 5k & 0.16675(67) & 0.27565(58) & \hspace{-0.30cm}$-$0.04609(22)  \\
0.24 & apeNEXT  & 5k & 0.14595(44) & 0.28866(36) & \hspace{-0.30cm}$-$0.04245(14)  \\
0.26 & apeNEXT  & 5k & 0.1267(12)  & 0.29949(56) & \hspace{-0.30cm}$-$0.03843(42)  \\
0.28 & apeNEXT  & 5k & 0.10607(61) & 0.30890(33) & \hspace{-0.30cm}$-$0.03353(17)  \\
0.30 & apeNEXT  & 5k & 0.08658(56) & 0.31653(30) & \hspace{-0.30cm}$-$0.02789(18)  \\
0.30 & APE100   & 5k & 0.08668(60) & 0.31639(35) & \hspace{-0.30cm}$-$0.02812(26)  \\
0.32 & APEmille & 5k & 0.06633(92) & 0.32255(41) & \hspace{-0.30cm}$-$0.02186(33)  \\
\hline
\hline
\end{tabular}
\label{summ_muimm_b145}
}

\TABLE[ht]{
\setlength{\tabcolsep}{0.9pc}
\centering \caption[]{Summary of the simulations at $\beta=1.30$ and
real chemical potential, $\hat\mu=\hat\mu_R$.}
\begin{tabular}{lcclll}
\hline
\hline
\multicolumn{1}{c}{$\hat\mu_R$} & machine & stat. & \multicolumn{1}{c}{$\langle L \rangle$}
& \multicolumn{1}{c}{$\langle\overline\psi\psi\rangle$}
& \multicolumn{1}{c}{$\Re \langle n_q\rangle$} \\
\hline
0.    & apeNEXT  & 5k    & 0.12667(23) & 0.36575(16)  &  0.00007(10) \\
0.05  & APEmille & 5k    & 0.12930(30) & 0.36421(25)  &  0.00958(21) \\
0.10  & apeNEXT  & 4k    & 0.13810(36) & 0.35908(25)  &  0.02086(11) \\
0.125 & apeNEXT  & 6k    & 0.14609(34) & 0.35416(20)  &  0.02819(10) \\
0.15  & apeNEXT  & 4.95k & 0.15507(30) & 0.34861(22)  &  0.03662(12) \\
0.175 & apeNEXT  & 6k    & 0.16648(29) & 0.34008(25)  &  0.04707(14) \\
0.20  & apeNEXT  & 6k    & 0.18064(55) & 0.32928(45)  &  0.06086(19) \\
0.225 & apeNEXT  & 5k    & 0.19691(35) & 0.31530(35)  &  0.07761(17) \\
0.25  & apeNEXT  & 4.8k  & 0.21636(46) & 0.29575(42)  &  0.09907(20) \\
0.275 & apeNEXT  & 6k    & 0.23737(30) & 0.27217(32)  &  0.12582(18) \\
0.30  & apeNEXT  & 5.75k & 0.25808(41) & 0.24655(49)  &  0.15608(26) \\
0.35  & apeNEXT  & 5.55k & 0.29092(26) & 0.20058(37)  &  0.22323(16) \\
0.40  & apeNEXT  & 5.3k  & 0.31534(30) & 0.16565(16)  &  0.29689(16) \\
0.45  & apeNEXT  & 5.25k & 0.33365(18) & 0.13838(11)  &  0.37795(16) \\
0.50  & apeNEXT  & 7.1k  & 0.34742(16) & 0.115763(67) &  0.46749(29) \\
0.55  & apeNEXT  & 4.8k  & 0.35841(16) & 0.096343(70) &  0.56615(21) \\
0.60  & apeNEXT  & 5.4k  & 0.36535(18) & 0.079910(73) &  0.67343(16) \\
0.65  & apeNEXT  & 5.25k & 0.36840(13) & 0.065787(51) &  0.78824(19) \\
0.70  & apeNEXT  & 5.4k  & 0.36749(15) & 0.053745(49) &  0.90967(16) \\
0.75  & apeNEXT  & 3.3k  & 0.36192(18) & 0.043309(57) &  1.03580(21) \\
0.80  & apeNEXT  & 6k    & 0.35154(11) & 0.034545(42) &  1.16451(15) \\
0.90  & apeNEXT  & 9k    & 0.31556(11) & 0.020801(39) &  1.41737(10) \\
1.00  & apeNEXT  & 7.95k & 0.26293(13) & 0.011606(28) &  1.63707(18) \\
1.10  & apeNEXT  & 6k    & 0.20260(22) & 0.006003(34) &  1.79870(14) \\
\hline
\hline
\end{tabular}
\label{summ_murea_b130}
}

\TABLE[ht]{
\setlength{\tabcolsep}{0.9pc}
\centering \caption[]{Summary of the simulations at $\beta=1.30$ and
imaginary chemical potential, $\hat\mu=i\hat\mu_I$.}
\begin{tabular}{lcclll}
\hline
\hline
\multicolumn{1}{c}{$\hat\mu_I$} & machine & stat. & \multicolumn{1}{c}{$\langle L \rangle$}
& \multicolumn{1}{c}{$\langle\overline\psi\psi\rangle$}
& \multicolumn{1}{c}{$\Im \langle n_q \rangle$} \\
\hline
0.   & apeNEXT & 5k &    0.12667(23) & 0.36575(16) &    0.00007(10)  \\
0.04 & apeNEXT & 5k &    0.12454(25) & 0.36652(18) & \hspace{-0.30cm}$-$0.00703(11)  \\
0.08 & apeNEXT & 5k &    0.11980(28) & 0.36908(15) & \hspace{-0.30cm}$-$0.013543(93) \\
0.12 & apeNEXT & 5k &    0.10978(34) & 0.37377(20) & \hspace{-0.30cm}$-$0.01802(10)  \\
0.16 & apeNEXT & 5k &    0.09860(26) & 0.37829(15) & \hspace{-0.30cm}$-$0.02113(11)  \\
0.20 & apeNEXT & 5k &    0.08489(27) & 0.38340(23) & \hspace{-0.30cm}$-$0.02162(18)  \\
0.24 & apeNEXT & 5k &    0.06919(31) & 0.38830(11) & \hspace{-0.30cm}$-$0.01977(12)  \\
0.28 & apeNEXT & 5k &    0.05219(29) & 0.39221(12) & \hspace{-0.30cm}$-$0.01640(17)  \\
0.32 & apeNEXT & 5k &    0.03386(23) & 0.39509(15) & \hspace{-0.30cm}$-$0.01123(14)  \\
0.36 & apeNEXT & 5k &    0.01534(27) & 0.39680(14) & \hspace{-0.30cm}$-$0.00523(17)  \\
0.40 & apeNEXT & 5k & \hspace{-0.30cm}$-$0.00373(36) & 0.39718(11) &    0.00120(14)  \\
0.44 & apeNEXT & 5k & \hspace{-0.30cm}$-$0.02215(21) & 0.396492(99)&    0.00757(15)  \\
0.48 & apeNEXT & 5k & \hspace{-0.30cm}$-$0.04075(23) & 0.39409(14) &    0.013311(94) \\
0.52 & apeNEXT & 5k & \hspace{-0.30cm}$-$0.05856(25) & 0.39085(13) &    0.01771(11)  \\
0.56 & apeNEXT & 5k & \hspace{-0.30cm}$-$0.07539(35) & 0.38647(13) &    0.02094(10)  \\
0.60 & apeNEXT & 5k & \hspace{-0.30cm}$-$0.08993(29) & 0.38160(14) &    0.02146(12)  \\
0.64 & apeNEXT & 5k & \hspace{-0.30cm}$-$0.10303(26) & 0.37660(17) &    0.02016(13)  \\
0.68 & apeNEXT & 5k & \hspace{-0.30cm}$-$0.11435(30) & 0.37165(20) &    0.016834(91) \\
0.72 & apeNEXT & 5k & \hspace{-0.30cm}$-$0.12147(30) & 0.36805(21) &    0.011327(91) \\
0.76 & apeNEXT & 5k & \hspace{-0.30cm}$-$0.12584(34) & 0.36612(19) &    0.004579(90) \\
0.80 & apeNEXT & 5k & \hspace{-0.30cm}$-$0.12630(27) & 0.36586(28) & \hspace{-0.30cm}$-$0.002841(88) \\
\hline
\hline
\end{tabular}
\label{summ_muimm_b130}
}

\TABLE[ht]{
\setlength{\tabcolsep}{0.75pc}
\centering \caption[]{Parameters of the interpolations of imaginary
chemical potential data at $\beta=1.90$.}
\begin{tabular}{cclllc}
\hline
\hline
observable & function &\multicolumn{1}{c}{$A$} & \multicolumn{1}{c}{$B$}
& \multicolumn{1}{c}{$C$} & $\chi^2$/d.o.f. \\
\hline
$\Im \langle n_q \rangle$ & Eq.~(\ref{poly_ndens})   & \hspace{-0.30cm}$-$0.37746(39) & 1.048(12)
& \hspace{-0.30cm}$-$0.138(81) & 1.11 \\
$\langle\overline\psi\psi\rangle$ & Eq.~(\ref{poly}) & 0.085780(17)   & 0.10687(77)  & 0.0592(61) & 0.55 \\
$\langle L \rangle$ & Eq.~(\ref{poly})               & 0.39706(12)    & \hspace{-0.30cm}$-$0.3511(57)
& \hspace{-0.30cm}$-$0.835(50) & 1.06 \\
\hline
$\Im \langle n_q \rangle$ & Eq.~(\ref{ratio_ndens})   & \hspace{-0.30cm}$-$0.37746(40) & 0.997(19)
& 0.136(80) & 1.11  \\
$\langle\overline\psi\psi\rangle$ & Eq.~(\ref{ratio}) & 0.085778(17)   & 0.0645(49)
& \hspace{-0.30cm}$-$0.497(50) & 0.58 \\
$\langle L \rangle$ & Eq.~(\ref{ratio})                & 0.39713(12)    & \hspace{-0.30cm}$-$1.029(30)
& \hspace{-0.30cm}$-$1.679(86) & 0.62 \\
\hline
\hline
\end{tabular}
\label{param_b190}
}

\TABLE[ht]{
\setlength{\tabcolsep}{0.75pc}
\centering \caption[]{Parameters of the interpolations of imaginary
chemical potential data at $\beta=1.45$. The last column gives the
largest value of $\hat\mu_I$ included in the fit.}
\begin{tabular}{cclllcc}
\hline
\hline
observable & function &\multicolumn{1}{c}{$A$} & \multicolumn{1}{c}{$B$}
& \multicolumn{1}{c}{$C$} & $\chi^2$/d.o.f. & $\bar\mu_I$ \\
\hline
$\Im \langle n_q \rangle$ & Eq.~(\ref{poly_ndens})   & \hspace{-0.30cm}$-$0.36670(97) & 2.708(93)
& 11.5(1.8) & 0.92 & 0.22 \\
$\langle\overline\psi\psi\rangle$ & Eq.~(\ref{poly}) & 0.20752(17)   & 1.202(22) & 3.72(40) & 1.16 & 0.24\\
$\langle L \rangle$ & Eq.~(\ref{poly})               & 0.25072(18)   & \hspace{-0.30cm}$-$1.200(23)
& \hspace{-0.30cm}$-$10.76(43) & 0.79 & 0.24 \\
\hline
$\Im \langle n_q \rangle$ & Eq.~(\ref{ratio_ndens})   & \hspace{-0.30cm}$-$0.36813(79) & 3.716(42)
& \hspace{-0.30cm}$-$2.20(27) & 1.33 & 0.24 \\
$\langle\overline\psi\psi\rangle$ & Eq.~(\ref{ratio}) & 0.20764(18)   & 0.42(10)
& \hspace{-0.30cm}$-$3.58(39) & 0.47 & 0.22 \\
$\langle L \rangle$ & Eq.~(\ref{ratio})               & 0.25083(18)   & \hspace{-0.30cm}$-$2.682(55)
& \hspace{-0.30cm}$-$5.68(30) & 0.74 & 0.22 \\
\hline
\hline
\end{tabular}
\label{param_b145}
}

\TABLE[ht]{
\setlength{\tabcolsep}{0.75pc}
\centering \caption[]{Parameters of the interpolations of imaginary
chemical potential data at $\beta=1.30$.}
\begin{tabular}{cclllc}
\hline
\hline
observable & function &\multicolumn{1}{c}{$A$} & \multicolumn{1}{c}{$B$}
& \multicolumn{1}{c}{$C$} & $\chi^2$/d.o.f. \\
\hline
$\Im \langle n_q \rangle$ & Eq.~(\ref{fourier_sin}) & \hspace{-0.30cm}$-$0.021582(37)
& \hspace{-0.30cm}$-$0.000611(35) &  & 1.25 \\
$\langle\overline\psi\psi\rangle$ & Eq.~(\ref{fourier_cos4}) & 0.38222(3)
& \hspace{-0.30cm}$-$0.015815(46) & \hspace{-0.30cm}$-$0.000769(44) & 0.67 \\
$\langle L \rangle$ & Eq.~(\ref{fourier_cos8}) & 0.12426(8)  & 0.002238(81) &  & 1.04 \\
\hline
\hline
\end{tabular}
\label{param_b130}
}

\end{document}